\documentclass[11pt,a4paper,fr]{smfart}

\usepackage{graphicx}

\font\tenbb=msbm10

\def\rR{\hbox{\tenbb R}}
\def\nN{\hbox{\tenbb N}}
\def\qQ{\hbox{\tenbb Q}}
\def\zZ{\hbox{\tenbb Z}}

\def\cal{\mathcal}

\def\esp{\vskip .6cm}
\def\pesp{\vskip .3cm}

\def\di{\displaystyle}

\newtheorem{thm}{Th\'eor\`eme}
\newtheorem{defi}{D\'efinition}
\newtheorem{lem}{Lemme}

\newtheorem{cor}{Corollaire}

\newcommand{\N}{\mathbb{N}}
\newcommand{\Z}{\mathbb{Z}}
\newcommand{\R}{\mathbb{R}}
\newcommand{\Q}{\mathbb{Q}}

\usepackage{graphicx} 

\begin{document}
\setcounter{tocdepth}{2}
\baselineskip 6mm

\title{Dynamique des nombres et physique des oscillateurs}

\author{Jacky CRESSON}
\address{Universit\'e de Franche-Comt\'e, Equipe de Math\'ematiques de Besan\c{c}on,
CNRS-UMR 6623, 16 route de Gray, 25030 Besan\c{c}on cedex, France.}

\maketitle

\tableofcontents

\section{Avant propos}

Le but de cet article est de pr\'esent\'e un exemple d'int\'eraction entre la th\'eorie des nombres
et la physique {\it exp\'erimentale}. J'insiste sur l'aspect exp\'erimental, car il n'est pas rare de trouver des
applications de la th\'eorie des nombres en physique th\'eorique, ou plus g\'en\'eralement en physique math\'ematique
(voir \cite{wald}). L'int\'er\^et est donc d'avoir sous la main un dispositif exp\'erimental qui permet de visualiser directement cette
int\'eraction.\\

Les math\'ematiques sous-jacente \`a cette int\'eraction sont \'el\'ementaires (essentiellement, elles tournent autour des
fractions continues). J'ai n\'eanmoins d\'ecid\'e d'en faire une pr\'esentation compl\`{e}te pour au moins deux raisons:\\

\begin{itemize}
\item La premi\`ere est que l'approche que nous allons emprunter n'est pas l'approche standard du sujet, et sugg\`ere des
interpr\'etations et des notions nouvelles.\\

\item La seconde est que ce texte est destin\'e \`a une communaut\'e beaucoup plus
large que les math\'ematiciens, qui n'est pas sp\'ecialement famili\`ere des notions math\'ematiques utilis\'ees.
\end{itemize}

\section{Introduction}

Les syst\`emes de communications n\'ecessitent la mise au point de circuits \'electroniques permettant de convertir, moduler
et d\'etecter des fr\'equences. Par exemple, la propagation des ondes radio est plus efficace pour des fr\'equences \'elev\'ees.
On cherche donc \`a transformer le signal afin de faire porter l'information initiale par un signal haute fr\'equence. Ce
proc\'ed\'e a de plus l'avantage de r\'eduire la taille des antennes n\'ecessaires \`a la reception (voir \cite{smith},p.487). La {\it boucle ouverte} est le composant \'electronique de base le plus r\'ependu pour
effectuer des modification de fr\'equences. Il est fond\'e sur un {\it m\'elangeur} qui th\'eoriquement ``fait" le produit
de deux signaux.\\

R\'ecemment, une s\'erie d'exp\'eriences destin\'ees \`a \'etudier le {\it bruit en 1/f}
, men\'ees par michel Planat au LPMO, ont conduit \`a un renouvellement de notre
comprehension du m\'elangeur et de la boucle ouverte. Ce renouvellement est d\^u en
partie \`a la grande pr\'ecision des mesures du spectre des fr\'equences et d'amplitudes
du signal de sortie.\\

Les principales nouveaut\'es dans l'analyse du m\'elangeur et de la boucle ouverte sont
les suivantes :

\begin{itemize}
\item Le spectre des fr\'equences est gouvern\'e par une analyse de type {\it diophantienne}.

\item Il existe une {\it r\'esolution} minimale (en temps et espace), intrins\`{e}que au syst\`{e}me, structurant
l'espace des fr\'equences via l'analyse diophantienne du point pr\'ec\'edent.
\end{itemize}

Le r\'esultat principal de cet article est un th\'eor\`eme abstrait permettant de pr\'edire la structure du spectre exp\'erimentale
de fr\'equences.\\

Les points 1 et 2 demandent l'introduction d'{\it espaces de r\'esolution} (appel\'es espaces de r\'esolution
{\it arithm\'etiques} dans \cite{CrDe}). Ils prennent en compte l'aspect diophantiens et les contraintes de
r\'esolution minimale en espace et temps.\\

La n\'ecessit\'e d'avoir une information sur l'approximation diophantienne des nombres r\'eels conduit
naturellement aux {\it fractions continues}. On en donne une pr\'e\-sen\-ta\-tion originale via les deux
op\'erations \'el\'ementaires $x\mapsto x+1$ et $x\mapsto 1/x$. Notamment, on obtient une
repr\'esentation nouvelle, \`a notre connaissance, de l'{\it arbre de Farey}. Ce choix
de construction et de repr\'esentation des nombres est dict\'e par la n\'ecessit\'e d'avoir une
traduction aussi simple que possible des contraintes de r\'esolution.\\

La contrainte de r\'esolution en {\it espace} s'interpr\'ete comme l'existence d'un {\it entier}, not\'e
$a_{max}$, au del\`{a} duquel, les nombres sont identifi\'es avec l'infini. On introduit ainsi une
structure d'{\it \'{e}chelle} naturelle, dans l'ensemble pr\'ec\'edent, en faisant appara\^{\i}tre des
z\^ones de {\it blocage} (ou d'{\it accrochage}) des nombres rationnels, des z\^ones de {\it transitions}
vers les z\^ones de blocage et enfin des z\^ones d'{\it instabilit\'e}, correspondant \`a des irrationnels.
Concr\'etement, on retrouve l'ensemble des fractions continues {\it \`a quotients partiels born\'es} par
$a_{max}$. On en donne une construction originale faisant intervenir un {\it syst\`{e}me dynamique} naturel
sur l'ensemble des fractions continues et conduisant \`a une {\it dynamique des nombres}. Cette
dynamique n'est apparente que lorsque $a_{max}$ est fini. On montre ainsi qu'il existe, d\`es qu'une
contrainte de r\'esolution est fix\'ee, une {\it hi\'erarchie} naturelle des nombres, hi\'erarchie qui disparait
si on regarde $\rR$ tout entier.\\

La contrainte en {\it temps}, se traduit par l'existence d'une borne $n_{max}$ \`a la longueur des
fractions continues. Autrement dit, le syst\`eme ne peut pas ``descendre" dans le d\'eveloppement en
fraction continue d'un nombre ind\'efiniment. On introduit alors une notion de z\^one {\it floue},
qui repr\'esente des endroits ou l'{\it analyse} du syst\`eme ne donne aucune information, les nombres \`a analyser
ayant un d\'eveloppement en fraction continue trop grand. Autrement dit, le syst\`eme fait bien quelque
chose, mais il est {\it impossible} de savoir quoi.\\

Le spectre des amplitudes ne se laisse pas aussi facilement capturer. Il n'existe pas pour le moment un analogue du th\'eor\`eme
de structure.

\section{Spectre de fr\'equences exp\'erimental}

On pr\'esente le mod\`{e}le de m\'elangeur et de filtre passe-bas qui nous servira
dans le reste de l'article. Nous d\'ecrivons le spectre des fr\'equences exp\'erimental
obtenu. Nous formulons ensuite notre approche du spectre des fr\'equences et l'hypoth\`ese principale de ce travail,
\`a savoir que la boucle ouverte ``fait" de l'approximation diophantienne des fr\'equences du signal.

\subsection{La boucle ouverte}

La {\it boucle ouverte} ou montage {\it superh\'et\'erodyne} d\'ecouvert par Armstrong et Schottky en 1924 permet
d'\'etudier un signal, appel\'e {\it signal de r\'ef\'erence} et not\'e $s_0 (t)$, \`a partir d'un signal connu
not\'e $s_1 (t)$. La fr\'equence de l'oscillateur de r\'ef\'erence est not\'ee $f_0 (t)$. Le signal connu est produit par
un oscillateur dit local de fr\'equence $f_1$.\\

La boucle ouverte est compos\'ee d'un {\it m\'elangeur} qui doit multiplier les deux signaux et d'un filtre dit passe-bas de
fr\'equence de coupure $f_c$, qui doit couper les fr\'equences au dessus de $f_c$. On a donc le dispositif suivant:
\begin{center}
\includegraphics[width=0.5\textwidth]{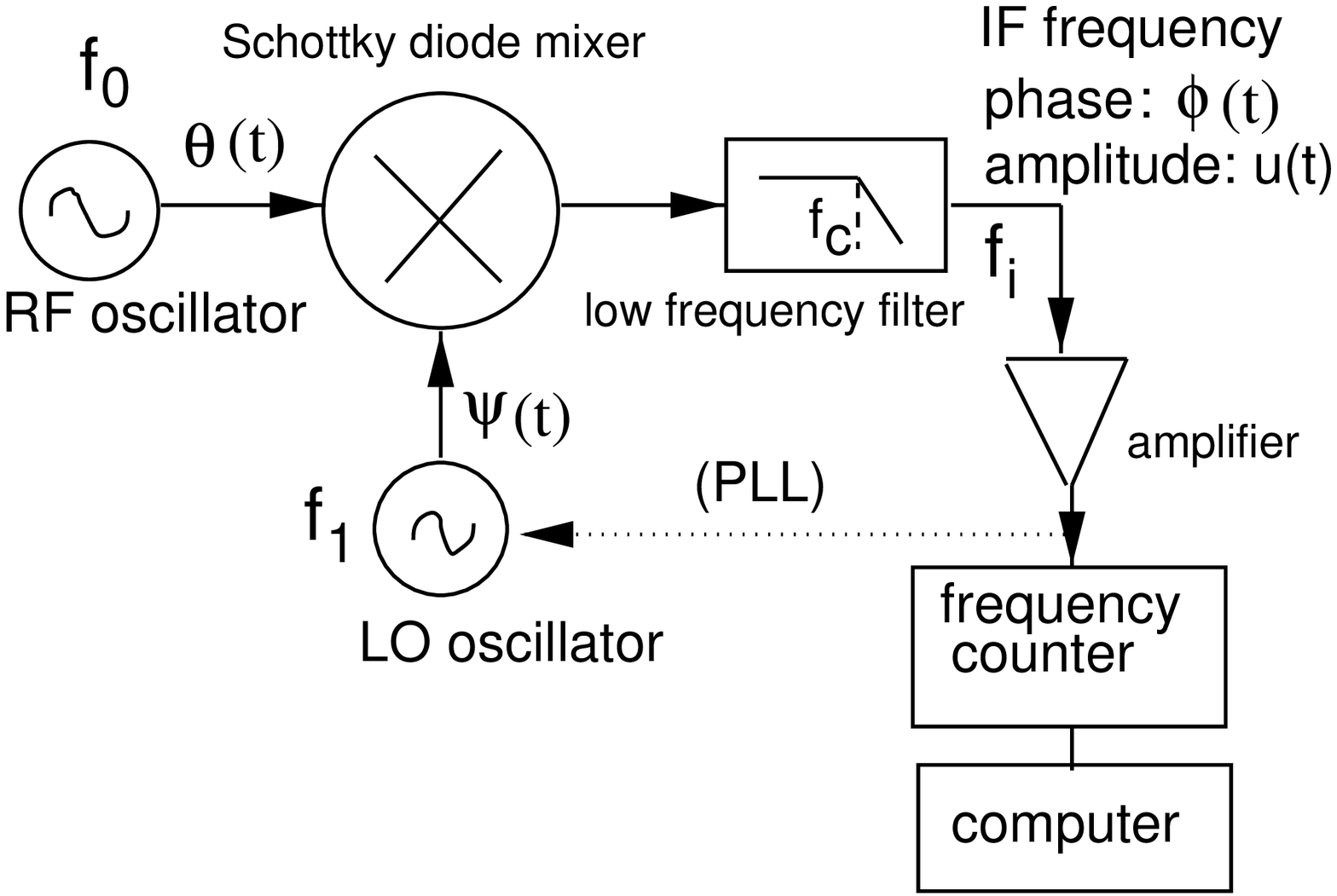}

{La boucle ouverte}
\end{center}

Que fait ce montage ?\\

Si l'on suppose que le m\'elangeur effectue r\'eellement le produit des deux signaux, nous obtenons en sortie du m\'elangeur
un signal de la forme
\begin{equation}
s(t)=\di {a_0 (t) a_1 (t) \over 2} \left ( \cos ((f_0 (t)+f_1 )t) +\cos ((f_0 (t)-f_1 (t))t) \right ) .
\end{equation}
Supposons que $f_0 (t)+f_1 >f_c$ pour tout $t$, on a en appliquant le filtre passe-bas un signal de la forme
\begin{equation}
s(t)=\di {a_0 (t) a_1 (t) \over 2} \cos ((f_0 (t)-f_1 (t))t) .
\end{equation}
On voit que l'action du m\'elangeur id\'eal est lin\'eaire en les fr\'equences et non lin\'eaire en les amplitudes.\\

Malheureusement, un m\'elangeur r\'eel a un comportement beaucoup plus compliqu\'e. Par ailleurs, la mod\'elisation des
m\'elangeurs est loin d'\^etre facile, m\^eme si par exemple, on connait exactement tous les composants \'electroniques
qui le constitue.
On renvoie \`a (\cite{lee},chapitre 12) pour plus de d\'etails. Le m\'elangeur fait en g\'en\'eral appara\^\i tre un
spectre dit d'intermodulation (\cite{lee},p.314), i.e. l'ensemble des combinaisons \`a coefficients entiers entre $f_0$
et $f_1$:
\begin{equation}
pf_0 -qf_1 ,\ \ p,q\in \zZ .
\end{equation}
La structure du spectre des fr\'equences obtenu en sortie de la boucle ouverte refl\`ete l'existence de ces modulations.
La mod\'elisation des composantes \'etant difficile il ne reste qu'une approche directe pour tenter de pr\'edire et expliquer
la structure du spectre des fr\'equences.

\subsection{R\'esultats exp\'{e}rimentaux}

Le spectre des fr\'equences est de la forme suivante:\\

\begin{center}
\includegraphics[width=0.5\textwidth]{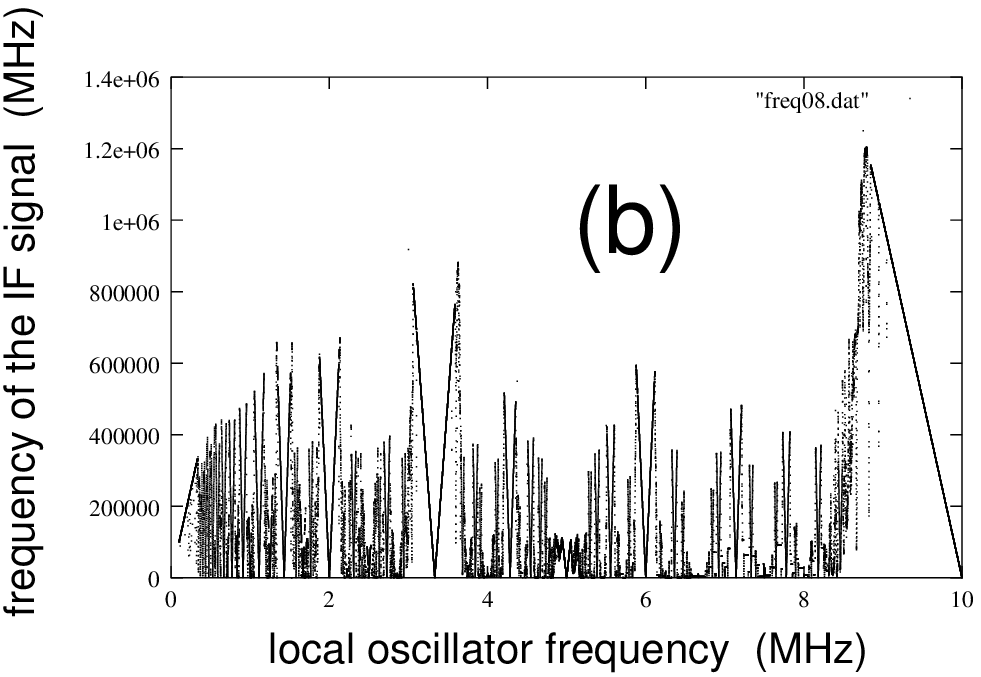}

{Spectre des fr\'equences}
\end{center}

Les principaux traits de sa structure sont:
\begin{itemize}
\item Les fr\'equences sont situ\'ees dans des bassins autour de fr\'equences rationnelles $p/q$, appell\'ees {\it
z\^ones d'accrochage}. Les bords
 de ces z\^ones sont not\'ees $\nu^- (p/q)$ et $\nu^+ (p/q)$.

\item Les bassins ne sont pas {\it sym\'etriques}. En effet, on observe que
\begin{equation}
p/q -\nu^- (p/q) \not= \nu^+ (p/q) -p/q .
\end{equation}
\end{itemize}

Dans la suite, nous allons d\'evelopper une th\'eorie {\it quantitative} qui rend compte de ces deux faits de mani\`ere pr\'ecise.

\section{Spectre de fr\'equence th\'eorique}

\subsection{Formalisation et hypoth\`ese diophantienne}

La principale diff\'erence entre le m\'elangeur id\'eal et le m\'elangeur ``r\'eel" est l'apparition d'harmoniques de la
forme
\begin{equation}
f_{p,q} (t)=pf_1 -qf_0 (t),
\end{equation}
avec $(p,q)\in \zZ^2 \setminus \{ (0,0)\}$.\\

L'action du filtre passe-bas de fr\'equence de coupure $f_c$ conduit \`a ne conserver que les harmoniques satisfaisant la
relation
\begin{equation}
\mid f_{p,q} (t) \mid < f_c .
\end{equation}
Notons
\begin{equation}
\nu (t)={f_0 (t) \over f_1} ,
\end{equation}
la fr\'equence {\it normalis\'ee}.\\

Le spectre des fr\'equences est donc gouvern\'e par une \'equation du type
\begin{equation}
\label{inifreq}
\left |
\nu (t) -\di {p\over q}
\right |
\leq \di {f_c \over q} .
\end{equation}
Comprendre le spectre des fr\'equences, c'est d\'eterminer les fr\'equences normalis\'ees autoris\'ees par la relation
(\ref{inifreq}).\\

Nous allons pour un moment oublier l'aspect temporel (donc la dynamique de $\nu (t)$) et nous concentrer sur l'aspect
statique du spectre des fr\'equences.\\

La premi\`ere id\'ee est que cette \'equation suffit \`a elle seule, \`a reconstruire le spectre des fr\'equences, ce qui est
en soit un peu os\'e, car cela suppose une ind\'ependance du spectre des fr\'equences vis \`a vis des amplitudes.\\

Sans hypoth\`ese sur la nature des approximations de $\nu$, on s'attend \`a trouver un bassin autour de chaque rationnel,
bord\'e par deux segments de pente $\pm q$. C'est effectivement le cas. Malheureusement, cette approche ne permet pas de
rendre compte de la {\it dissym\'etrie} des bassins observ\'ee exp\'erimentalement.\\

Si l'\'equation (\ref{inifreq}) contient l'essentiel de l'information
sur la nature du spectre de fr\'equence, c'est donc que {\it l'approximation de $\nu$ n'est pas triviale}. Il reste donc
\`a d\'eterminer la {\it nature} de l'approximation effectu\'ee par le d\'etecteur.\\

L'hypoth\`ese que nous allons faire est que le d\'etecteur a un comportement {\it diophantien}, i.e. que les
approximations d'une fr\'equences $\nu$ sont effectu\'ees par des {\it convergents}. Autrement dit, on doit v\'erifier
la condition
$$
\left |
\nu (t) -\di {p_i\over q_i}
\right |
\leq \di {f_c \over f_0 q_i} \leq {1\over \nu_{i+1} q_i^2 },
\eqno{(D)}
$$
o\`u $\nu=[\nu_0 ,\nu_1 ,\dots ]$ repr\'esente le d\'eveloppement en fraction continue de $\nu$, et
$p_i/q_i =[\nu_0 ,\dots ,\nu_i ]$ est le $i$-\`eme convergent de $\nu$.

\subsection{Principaux r\'esultats}

Cette conditions a plusieurs cons\'equences, qu'il conviendra ensuite de v\'erifier exp\'erimentalement:\\

i) Les fr\'equences $\nu$ observ\'ees en sortie du d\'etecteur sont tr\`es contraintes par (D). Soit $p/q$ un
rationnel fix\'e, alors les $\nu$ associ\'es ont un d\'eveloppement en fraction continue qui v\'erifie
\begin{equation}
\label{amax}
1\leq \nu_{i+1} <\di {f_1 \over f_c q} .
\end{equation}

ii) La condition (\ref{amax}) impose un seuil maximal pour $q$, \`a savoir
\begin{equation}
q\leq \left [
\di {f_1 \over f_0 }
\right ]
.
\end{equation}
Comme le d\'enominateur $q_i$ d'un convergent croit avec $i$, cela impose une profondeur maximale dans le
d\'eveloppement en fraction continue de $\nu$.\\

Nous allons r\'esumer la discussion pr\'ec\'edente par le th\'eor\`eme suivant, qui pour un rationnel donn\'e $p/q$, d\'ecrit
l'ensemble des nombres r\'eels {\it admissibles} sous la contrainte (\ref{inifreq}) et l'hypoth\`ese diophantienne.

\begin{thm}
\label{spectre static}
Soit $p/q$ une fraction irr\'eductible. On note $S(p/q)$ l'ensemble des nombres r\'eels satisfaisant (\ref{inifreq}) sous
l'hyppoth\`ese diophantienne (D). Alors, on a:\\

- L'ensemble $S(p/q)$ est non vide si et seulement si $q\leq \left [ f_1 /f_c \right ]$, o\`u $[x]$ d\'esigne la partie enti\`ere
de $x$.\\

- Soit $[a_1 ,\dots ,a_n]$ le d\'eveloppement en fractions continues de $p/q$. L'ensemble $S(p/q)$ est l'ensemble des
nombres r\'eesl $x\in \rR$ de d\'eveloppement en fractions continues $[x_1,\dots ,x_k ,\dots ]$ tels que
\begin{equation}
x_i =a_i \ \mbox{\rm pour}\ i=1,\dots, n,
\end{equation}
et
\begin{equation}
x_{n+1} \leq f_1 /f_c q .
\end{equation}
\end{thm}

Nous appelons {\it spectre th\'eorique} des fr\'equences l'ensemble
\begin{equation}
S_{f_1 /f_c} =\left \{
\nu \in \rR \ \mbox{\rm satisfaisant}\ (D) \right \} .
\end{equation}
Le th\'eor\`eme \ref{spectre static} permet de pr\'eciser la structure de $S_{f_1 /f_c}$. C'est un th\'eor\`eme de nature
pr\'edictive puisqu'il permet de reconstruire le spectre des fr\'equences \`a partir de la donn\'ee de $f_1$ et $f_c$. Afin
de comparer les pr\'edictions de notre th\'eorie avec le spectre exp\'erimental ${\cal S}_{f_1 /f_c}$, nous d\'emontrons
le r\'esultat suivant:

\begin{lem}
\label{bordaccro}
Soit $p/q$ une fraction irr\'eductible , avec $q\leq \left [ f_1 /f_c \right ]$ et $[a_0 ,a_1 ,\dots ,a_n ]$ son
d\'eveloppement en fraction continue. Le bord de la z\^one d'accrochage est donn\'e par
\begin{equation}
\label{formeaccro}
\left .
\begin{array}{lll}
\nu^{\sigma } & = & [a_0 ,\dots ,a_n ,a] ,  \\
\nu^{-\sigma } & = & [a_0 ,\dots ,a_n -1 ,1,a ] ,
\end{array}
\right .
\end{equation}
avec $a=[ f_1 /f_c q ]$ et $\sigma =+$ si $n$ est pair et $\sigma =-$ si $n$ est impair.
\end{lem}

La d\'emonstration est donn\'ee \`a la section \ref{zoneaccro}.\\

Nous avons maintenant la caract\'erisation analytique des principaux \'el\'ements g\'eom\'etriques du spectre des fr\'equences
th\'eoriques.

\subsection{Confirmation exp\'{e}rimentale}

On peut tester la validit\'e de l'hypoth\`ese diophantienne (D) via le th\'eor\`eme
\ref{spectre static} et le lemme \ref{bordaccro}.

\begin{itemize}
\item Michel Planat et Serge Dos Santos \cite{serge} ont montr\'e que les bords des z\^ones d'accrochage sont de la forme
(\ref{formeaccro}).\\

\item L'hypoth\`ese diophantienne pr\'edit l'existence dans une z\^one d'accrochage donn\'ee,
d'une borne sup\'erieure pour les quotients partiels $a_{\rm max}$. Cette borne se retrouve dans l'expression des bords de la z\^one
d'accrochage. Michel Planat et Jean-Philippe Marillet \cite{mar} ont montr\'e que $a_{\rm max}$ est bien de la forme
$f_1 /f_c q$.
\end{itemize}

\subsection{Exploration de la condition diophantienne}

Pour mieux comprendre la nature de l'hypoth\`ese diophantienne (D), nous allons proc\'eder en deux \'etapes:

\begin{itemize}
\item La premi\`ere consiste \`a comprendre l'effet d'une contrainte, que j'appellerai de {\it r\'esolution}, sur les quotients
partiels d'une fraction continue, qui d\'ecoule du point i) ci-dessus. Autrement dit, nous allons regarder la structure des fractions
continues \`a quotients partiels born\'es par un $a_{\rm max}\in \nN$ fix\'e. Cette \'etude, faite dans la prochaine section sera
riche d'enseignement sur la diff\'erence essentielle qu'il existe entre travailler sur les nombres r\'eels, et travailler
sous une contrainte de r\'esolution. En particulier, nous verrons que l'existence d'une r\'esolution induit de fait une
{\it hi\'erarchie} des nombres et donne naissance \`a une {\it dynamique des nombres}.\\

\item La seconde \'etape prend en compte l'aspect dynamique du spectre des fr\'equences. La d\'etection dans un bassin donn\'e fait
appara\^\i tre des sauts de fr\'equences. On donne ici une justification th\'eorique de l'existence de ces sauts qui fait intervenir
de mani\`ere essentielle l'hypoth\`ese diophantienne.
\end{itemize}

\section{Espaces de r\'esolution: aspects g\'eom\'etriques}

Ce paragraphe donne une construction  g\'eom\'etrique de l'ensemble des fractions continues \`a quotient partiels
born\'es faisant appara\^\i tre une structure d'arbre. Cette construction n'est sans doute pas nouvelle, mais nous n'avons
pas trouv\'e de r\'ef\'erence faisant appara\^\i tre simplement les structures dont nous avons besoin. On renvoie au livre de G.H. Hardy et E.M. Wright (\cite{HW},p.
164-169) pour la pr\'esentation standard.

\subsection{G\'eom\'etrie des fractions continues}

\subsubsection{Repr\'esentation des fractions irr\'eductibles}

Soit $p/q$ une fraction {\it
irr\'eductible} de $\qQ$. On lui associe le point $(q,p)\in
\zZ^2$, ou de mani\`ere \'equivalente, la droite de $\zZ^2$
passant par $0$ et $(q,p)$, de pente $p/q$ et d'\'equation
$qx-py=0$.

On a donc une bijection entre $\qQ\bigcup \{ \infty \}$ et $P^1 (\zZ^2 )$, l'ensemble des droites vectorielles de
$\zZ^2$, d\'efinie comme l'ensemble des points de $\zZ^2$ modulo l'\'equivalence $(q,p )\sim (q' ,p')$
si et seulement si il existe un entier $\lambda\in \zZ$ tel que $(q,p)=\lambda (q',p')$ ou $(q',p')=\lambda
(q,p)$.

\begin{figure}
\label{irred}
\centering
\includegraphics[width=1\textwidth]{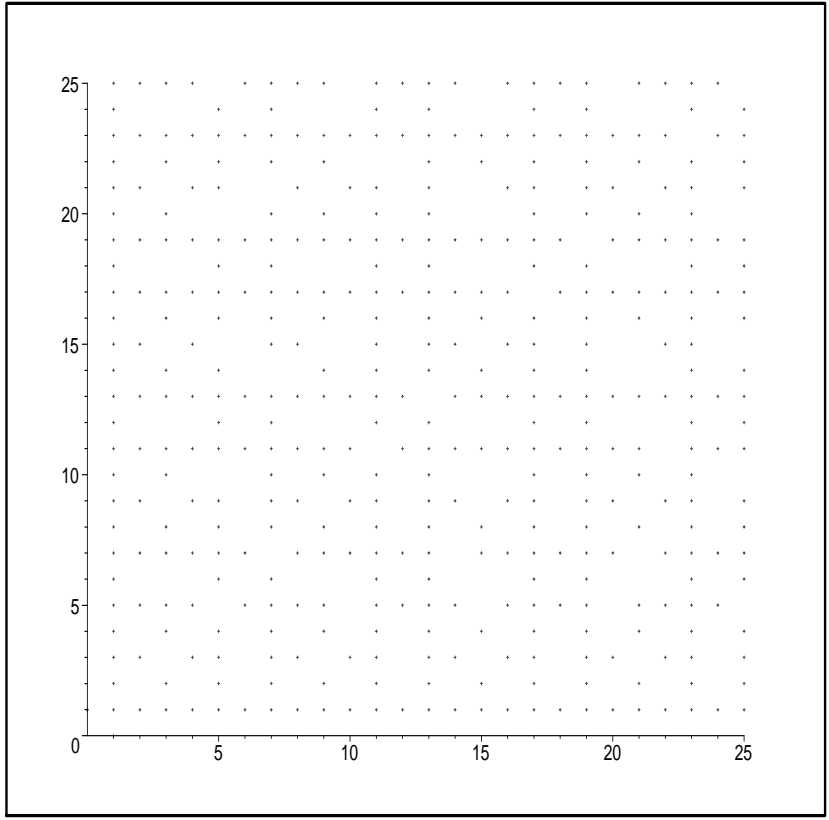}
\caption{Points premiers de $Z^2$ ($pgcd(p,q)=1$).}
\end{figure}

Chaque droite $D$ de $P^1 (\zZ^2 )$ est isomorphe \`a $\zZ$, et est engendr\'e par un des deux points
$(q,p)$, $(-q,-p)$ de $D$ v\'erifiant $<q,p>=1$. Ces points sont dit {\it premiers} dans $\zZ^2$.
On note $\cal P$ l'ensemble des points premiers de $\zZ^2$.
\pesp
L'anneau des $\zZ$-matrices $2\times 2$, not\'e $M_2 (\zZ )$, agit
naturellement sur $\zZ^2$ :
\begin{equation}
\left .
\begin{array}{llll}
\forall A=\left (
\begin{array}{lll}
a & b \\
c & d
\end{array}
\right )
\in M_2 (\zZ ) , &  \zZ^2 & \stackrel{A}{\rightarrow} & \zZ^2 , \\
 & (q,p) & \mapsto & (aq+bp,cq+dp) .
\end{array}
\right .
\end{equation}
Cette action induit une action sur $\qQ$ via les {\it transformations de
M\"obius} :
\begin{equation}
\left .
\begin{array}{llll}
A\in M_2 (\zZ ) , & \qQ & \stackrel{A}{\rightarrow} & \qQ , \\
 & z=p/q & \mapsto & (cz+d)/(az+b) .
\end{array}
\right .
\end{equation}
La matrice $A$ pr\'eserve $\cal P$ si et seulement si $\mid \mbox{\rm det} (A)\mid =1$, i.e. $A$ est inversible dans
$M_2 (\zZ )$. On consid\`ere donc l'action de $GL_2 (\zZ )$, l'ensemble des matrices inversibles de $M_2 (\zZ )$, sur
$\qQ$, via les transformations de M\"obius.

\subsubsection{Fractions continues et $F_2^+$}

On renvoie \`a Khintchine \cite{Kh} pour plus de d\'etails.
\esp
Soient $(a_0 ,\dots ,a_n )$ une suite finie d'entiers avec $a_n \not= 0$. On note
$[a_0 ,\dots ,a_n ]$ la fraction continue finie
\begin{equation}
a_0 +\di {1\over a_1 +\di {1\over a_2 + \di {1\over \dots +\di {1\over a_n}}}} .
\end{equation}
On conservera la m\^eme notation pour une suite de longueur infinie.

Supposons tous les $a_i >0$ pour $i>0$, alors tout {\it irrationnel} a une {\it unique}
repr\'esentation comme fraction continue infinie. Par contre, l'\'egalit\'e
\begin{equation}
\label{nonun}
[a_0 ,\dots ,a_n ] =[a_0 ,\dots ,a_n -1 ,1] ,
\end{equation}
montre qu'un rationnel poss\`{e}de deux \'ecritures. On en d\'eduit deux fa\c{c}on de rendre unique la
repr\'esentation d'un rationnel :
\pesp
i - tout nombre rationnel poss\`{e}de un unique d\'eveloppement en fraction continue de longueur pair (ou
impair).
\pesp
ii - tout nombre rationnel poss\`{e}de un unique d\'eveloppement en fraction continue se terminant par
un entier $>1$.
\pesp
La premi\`ere repr\'esentation est adapt\'ee \`a l'introduction du {\it groupe modulaire}. La
seconde supprime les extensions virtuelles de la fraction continue via (\ref{nonun}). Elle est bien
adapt\'ee \`a la construction de l'espace de r\'esolution.
\pesp
D'apr\`es l'algorithme des fractions continues, il est possible de
construire toutes les fractions continues via les applications \'el\'ementaires de
{\it translation}, not\'ee $T$, et d'{\it inversion}, not\'ee $S$ :
\begin{equation}
T\ :\
\left .
\begin{array}{lll}
\rR & \rightarrow & \rR ,\\
x & \mapsto x+1 ,
\end{array}
\right .
\ \ \
\mbox{\rm et}
\ \ \
S\ :\
\left .
\begin{array}{lll}
\rR & \rightarrow & \rR ,\\
x & \mapsto 1/x .
\end{array}
\right .
\end{equation}
On peut restreindre l'action de $T$ (resp. $S$) \`a $Q$. Dans ce cas, on a deux homographies qui sont repr\'esent\'ees
dans $M_2 (\zZ )$ par les matrices
\begin{equation}
T=
\left (
\begin{array}{ll}
1 & 0 \\
1 & 1
\end{array}
\right )
, \ \ \ \mbox{\rm et}\ \ \
S=
\left (
\begin{array}{ll}
0 & 1 \\
1 & 0
\end{array}
\right )
.
\end{equation}
Une fraction continue $[a_0 ,\dots ,a_n ]$ s'\'ecrit donc $T^{a_0} S T^{a_1} S \dots S T^{a_n } (1,0).$ Pour
obtenir une repr\'esentation unique, on choisi des repr\'esentations de longueur paire. On introduit la
matrice $J =STS$, de la forme
\begin{equation}
J=
\left (
\begin{array}{ll}
1 & 1 \\
0 & 1
\end{array}
\right )
,
\end{equation}
qui correspond \`a la transformation $x\rightarrow \di {x\over x+1}$ sur $\qQ$.\\

On a :

\begin{thm}
Tout nombre rationnel $[a_0 ,\dots ,a_{2n} ]$ admet une unique repr\'esentation de la forme
\begin{equation}
T^{a_0} J^{a_1} \dots J^{a_{2n-1}} T^{a_{2n}} (1,0) .
\end{equation}
\end{thm}

Les matrices $T$ et $J$ sont {\it unimodulaires} (de d\'eterminant $1$). Elles engendrent le {\it groupe
modulaire} $PSL_2 (\zZ )$, qui est isomorphe au groupe libre de rang 2 $F_2$, $T$ et $J$
\'etant deux g\'en\'erateurs libres. On note $F_2^+$ le semi-groupe des mots \'ecrit avec des
puissances positives de $T$ et $J$. On a

\begin{cor}
L'application
\begin{equation}
\left .
\begin{array}{lll}
\qQ & \rightarrow & F_2^+ , \\
\left [ a_0 ,\dots ,a_{2n} \right ] & \mapsto & T^{a_0} J^{a_1} \dots J^{a_{2n-1}} T^{a_{2n}} ,
\end{array}
\right .
\end{equation}
est une bijection. Le groupe libre $F_2^+$ agit \`a gauche sur $\qQ$.
\end{cor}

La d\'emonstration d\'ecoule du th\'eor\`eme pr\'ec\'edent.

\subsection{L'arbre de Farey}

\subsubsection{Terminologie sur les arbres}

On renvoie au livre de Serre (\cite{Se2}, $\S$.2.2, p.28) pour plus de d\'etails. On
rappelle qu'un {\it arbre} est un graphe connexe, non vide, sans circuit. On adopte la convention
suivante sur la repr\'esentation d'un arbre par un dessin : un point correspond \`a un {\it sommet} de l'arbre, et une
ligne joignant deux points marqu\'es correspond \`a une {\it ar\^etes}. Si l'arbre est orient\'e, une
ar\^ete $\{ P,Q\}$ \'etant donn\'e, on appelle le sommet $P$, l'{\it origine} de $\{ P,Q\}$ et
$Q$ le sommet {\it terminal} de $\{ P,Q\}$. Ces deux sommets sont les {\it extr\'emit\'es} de $\{ P,Q\}$.

Un sommet $P$ d'un arbre $\Gamma$ orient\'e \'etant donn\'e, on appellera {\it fils} de $P$ l'ensemble
des sommets terminaux des ar\^etes ayant $P$ comme origine. On appellera {\it p\`ere} de $P$, l'origine
de l'ar\^ete ayant $P$ comme sommet terminal.

\subsubsection{Arbre de Farey}

Habituellement, on repr\'esente l'arbre de Farey via l'action du groupe
modulaire sur le demi-plan de Poincar\'e (ou de mani\`ere \'equivalente sur le disque de
Poincar\'e). On en donne ici une repr\'esentation dans $\zZ^2$, plus commode pour la suite.
\pesp
On note $L_{\infty}$ la droite passant par $(1,0)$, engendr\'e par l'action de $T$ sur le segment
$[(1,0),(1,1)]$. Elle est de pente $\infty$.

De m\^eme, on note $L_0$ la droite passant par $(0,1)$, engendr\'e par l'action de $J$ sur le segment
$[(0,1),(1,1)]$. Elle est de pente $0$.
\pesp
L'action de $F_2^+$ sur $\zZ^2$, induit une action de $F_2^+$ sur $L_0$ et $L_{\infty}$. On obtient la
figure suivante :

\begin{figure}[h]
\label{PFTree}
\centering
\includegraphics[width=1\textwidth]{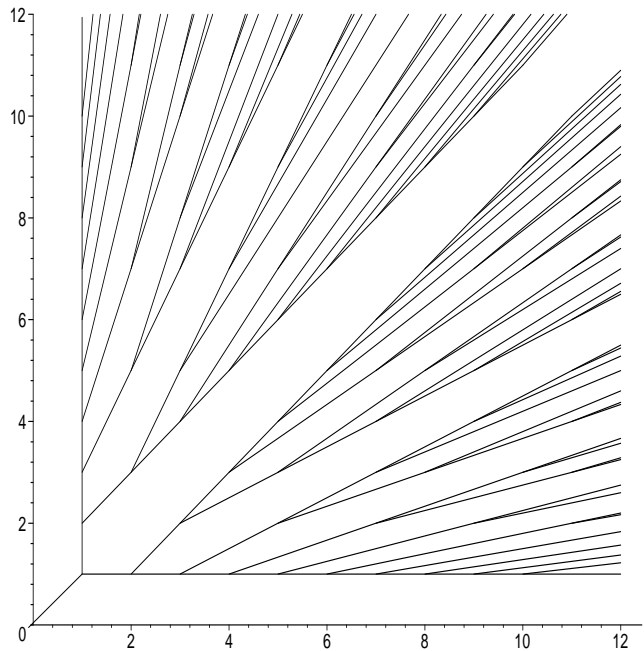}
\caption{L'arbre de Farey}
\end{figure}

On note $\cal T$ l'ensemble ainsi obtenu. On a :

\begin{thm}
L'ensemble $\cal T$ est un arbre (ou plut\^{o}t, la r\'{e}alisation geom\'etrique d'un arbre) dont les
sommets sont les points $(q,p)\in\zZ^2$ irr\'eductibles.
\end{thm}

Ce r\'esultat est classique (au moins dans le demi-plan de Poincar\'e, voir (\cite{Se2},$\S$.4.2,p.52-53)).
\pesp
Nous allons pr\'eciser, la relation entre le d\'eveloppement en fraction continue d'un sommet de $\cal T$,
celui de ses fils et de son p\`ere. Pour cela, nous introduisons la notion de {\it branches} et {\it rameaux}
de l'arbre $\cal T$.

\begin{defi}
Une branche de $\cal T$ est l'image par un mot de $F_2^+$ des droites $L_0$ ou $L_{\infty}$. Soit $B$ une
branche de $\cal T$, on appellera rameau de $B$ en $P$, une branche distincte de $B$ ayant pour origine le
sommet $P$.
\end{defi}

Une cons\'equence du th\'{e}or\`{e}me pr\'ec\'edent est :

\begin{cor}
Soit $M=(q,p)\in \zZ^2$, $(q,p)\not= (1,1)$, un sommet de $\cal T$, alors $M$ appartiens \`a deux branches distinctes
$B_M^m$ et $B_M^f$, appel\'ees branche m\`ere et fille. La branche m\`ere admet la branche fille comme rameau en $M$.
\end{cor}

La caract\'erisation de l'arbre $\cal T$ en terme de fractions continues s'\'enonce maintenant comme suit :

\begin{thm}
Soit $M=(q,p)$ un sommet de $\cal T$, tel que $p/q=[a_0 ,\dots a_{2n} ]$. On note $B_M^m$ et
$B_M^f$ ses branches m\`ere et fille respectivement. On a :

i - l'origine de la branche m\`ere est
$$
\left .
\begin{array}{ll}
\left [ a_0 ,\dots ,a_{2n-1} ,1 \right ] & \mbox{\rm si}\ \ a_{2n} >1 , \\
\left [ a_0 ,\dots ,a_{2n-2} +1 \right ] & \mbox{\rm si}\ \  a_{2n} =1 .
\end{array}
\right .
$$
ii - la pente de la branche m\`ere est
$$
\left .
\begin{array}{ll}
\left [ 0,a_0 ,\dots ,a_{2n-1} -1,1 \right ] & \mbox{\rm si}\ \ a_{2n} >1 , \\
\left [ 0,a_0 ,\dots ,a_{2n-2} \right ] & \mbox{\rm si}\ \ a_{2n} =1 .
\end{array}
\right .
$$
iii - La pente de la branche fille est
$$
\left .
\begin{array}{ll}
\left [ a_0 ,\dots ,a_{2n} -1 \right ]\ & \mbox{\rm si}\ \ a_{2n} >1 , \\
\left [ a_0 ,\dots ,a_{2n-1} \right ] & \mbox{\rm si}\ \ a_{2n} =1 .
\end{array}
\right .
$$
\end{thm}

\begin{proof}
Elle repose sur la construction it\'erative de l'arbre de Farey.
\pesp
i) La branche m\`ere de $M$ est l'image par un mot de $w \in F_2^+$ de la droite
$L_0$ ou $L_{\infty}$.

Supposons que $B_M^m$ soit l'image de $L_{\infty}$ par $w$ (le cas de $L_0$ se d\'emontre de la m\^eme
mani\`ere). L'origine de $B_M^m$ est donc $w(1,0)$. Si $M \not= w(1,0)$, il existe un entier $k>0$
tel que
\begin{equation}
\label{egalm}
M=w\di T^k (1,0)=T^{a_0 }\dots \dots  T^{a_{2n}} (1,0) ,
\end{equation}
ou $w\di T^k$ est le mot obtenu par concat\'enation de $w$ et $\di T^k$.

Comme le mot $w$ ne se termine pas par un $T^l$, $l>0$, on d\'eduit de (\ref{egalm}) et de l'unicit\'e
de l'\'ecriture de $M$, $k=\di a_{2n}$ et $w=T^{a_0} \dots J^{a_{2n-1}} T$ si $a_{2n} >1$, d'o\`u
l'origine de la branche m\`ere dans ce cas est $w(1,0)=[a_0 ,\dots ,a_{2n-1} ,1]$.

Si $a_{2n} =1$, on \'ecrit $[a_0 ,\dots ,a_{2n-1} ,1]=[a_0 ,\dots ,a_{2n-1} +1]$, d'o\`u
$M=w\di T^{a_{2n-1}} (1,0)$, avec $w=T^{a_0 } B^{a_1} \dots T^{a_{2n+2} +1}$.
On a donc l'origine de la branche m\`ere donn\'ee par $w(1,0)=[a_0 ,\dots ,a_{2n+2} +1]$.
\pesp
ii - Il suffit de noter que $L_0$ (resp. $L_{\infty}$) est parall\'ele \`a la droite passant par
$(0,0)$ et $(1,0)$ (resp. $(0,0)$ et $(0,1)$). Quel que soit le mot $w\in F_2^+$, on a $w.(0,0)=(0,0)$
car $w$ est une application lin\'{e}aire. En utilisant i), la pente de la branche m\`ere de $M=w.(1,0)$,
$w=T^{a_0 }\dots \dots  T^{a_{2n}}$ est donc donn\'ee par $T^0 J^{a_1} T^{a_2} \dots J^{a_{2n-1}} T^{1} (1,0)$
si $a_{2n} >1$ et par $T^0 J^{a_1} \dots , T^{a_{2n-2}} (1,0)$ si $a_{2n} =1$.
\pesp
iii - La d\'emonstration est analogue \`a ii) en consid\'erant la branche fille comme une branche m\`ere d'origine
$[a_0 ,\dots ,a_{2n}]$.
\end{proof}

\subsection{Ensembles de r\'esolution}

Le spectre des fr\'equences serait donn\'e par l'ensemble pr\'ec\'edent si aucune contrainte
de r\'esolution n'existait, i.e. dans un syst\`eme id\'eal (au sens math\'ematique). Les
r\'esultats exp\'erimentaux et la physique, imposent l'existence d'une r\'esolution minimale.
Dans ce paragraphe, nous interpr\'etons cette contrainte et en donnons l'effet sur le
spectre des fr\'equences.

\subsubsection{La contrainte de r\'esolution}

Il faut traduire la notion intuitive de {\it r\'esolution} de mani\`ere \`a en obtenir une traduction simple sur l'ensemble
des fractions continues.\\

\begin{itemize}
\item {\it Hypoth\`ese de r\'esolution (nombres)}. Soit $a>0$ un entier. On identifie tout nombre r\'eel $x\geq a$ \`a
$\infty$.
\end{itemize}

On remarque que cette hypoth\`ese de r\'esolution {\it \`a l'infinie} implique, via l'action de
l'application $S$, une condition de r\'esolution {\it en z\'ero}. En effet, tous les
nombres r\'eels $0\leq x \leq 1/a$ sont identifi\'es \`a $0$.
\pesp
On note ${\cal R}_a$ l'ensemble des nombres r\'eels obtenus.
\pesp
L' hypoth\`ese se traduit sur les mots {\it admissibles} de $F_2^+$.

\begin{itemize}
\item {\it Hypoth\`ese de r\'esolution (mots)}. Les seuls mots admissibles de $F_2^+$ sont ceux ne contenant
que des $T^i$ avec $i<a$.
\end{itemize}

On en d\'eduit donc le th\'eor\`eme suivant :

\begin{thm}
L'ensemble de r\'esolution ${\cal R}_a$ est l'ensemble des fractions continues \`a quotients
partiels born\'es.
\end{thm}

Ce th\'eor\`eme n'apporte pas beaucoup \`a la compr\'ehension de l'ensemble ${\cal R}_a$.
Nous allons pr\'eciser la structure g\'eom\'etrique et dynamique de cet ensemble dans le
prochain paragraphe.

\section{Espaces de r\'esolution: aspects dynamiques}

\subsection{Syst\`eme dynamique de r\'esolution}

On travaille maintenant dans $\bar{\rR}^+ =\rR^+ \cup \{ \infty \}$. Pour tout $a\in \N^*$, nous allons introduire
une application naturelle sur $\bar{\rR}$ appell\'ee {\it application de r\'esolution}. \\

Soit $a\in \nN^*$, on note
$F_2^+ (a)$ l'ensemble des mots de $F_2^+$ ne contenant pas de sous mots $T^k$ ou $J^k$ avec $k\geq a$.

\begin{defi}
Soit $a\in \nN^*$, $w$ un mot fini de $F_2^+$, $w=w_1 \dots w_n$, on d\'efinit l'application de
$F_2^+$ dans $F_2^+ $ qui \`a $w$ associe $w_a$ obtenu en rempla\c{c}ant le premier $T^i$ ou $J^i$ avec
$i\geq a$ par $\infty$ ou $O$ respectivement. On note $R_a$ cette application.
\end{defi}

L'application $R_a$ d\'efini un syst\`eme dynamique sur $F_2^+$. L'ensemble invariant maximal de $R_a$
est $F_2^+ (a)$. La traduction sur les nombres se fait via l'application
\begin{equation}
\left .
\begin{array}{llll}
r_a \ :\ & \bar{\rR}^+ & \rightarrow & \bar{\rR}^+ , \\
 & x=w(1,0) & \mapsto & x_a =R_a (w) (1,0) .
\end{array}
\right .
\end{equation}
L'application $r_a$ d\'efini un syst\`eme dynamique sur $\bar{\rR}^+$. Ce syst\`eme dynamique est \`a ma connaissance
nouveau. Son graphe est donn\'e pour $a=3$ par:

\begin{figure}[h]
\label{esc201}
\centering
\includegraphics[width=1\textwidth]{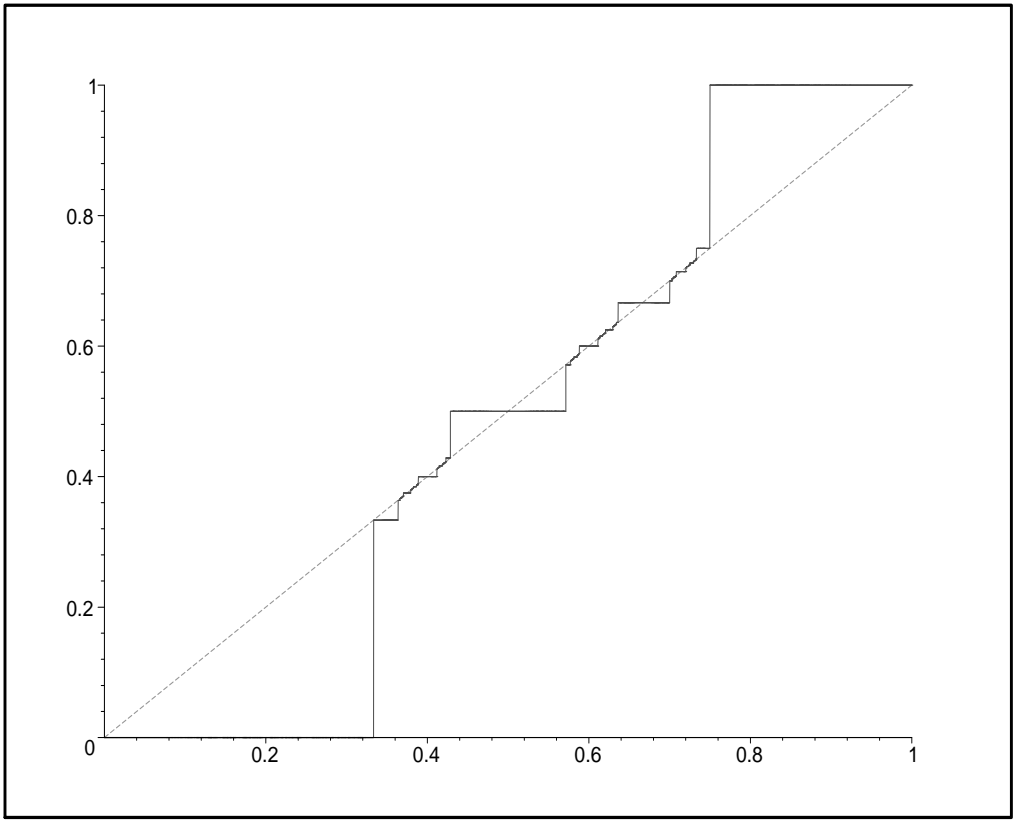}
\caption{Graphe de l'application $r_3$.}
\end{figure}

L'ensemble invariant maximal de $r_a$ est ${\cal R}_a$.\\

Cette vision dynamique de l'ensemble des fractions continues \`a quotients partiels born\'es permet de d\'efinir une
dynamique {\it naturelle} des nombres. Pr\'ecisons tout d'abord la structure g\'eom\'etrique de ${\cal R}_a$.

\subsection{Arbre de r\'esolution}

Avant de formuler le th\'eor\`eme de structure sur ${\cal R}_a$, nous pouvons, en utilisant le proc\'ed\'e de construction
utilis\'e pour l'arbre de Farey, construire l'ensemble ${\cal R}_a$ pour un $a$ fix\'e. Par exemple, dans le cas $a=3$,
on a:

\begin{figure}[h]
\label{tructree01}
\centering
\includegraphics[width=1\textwidth]{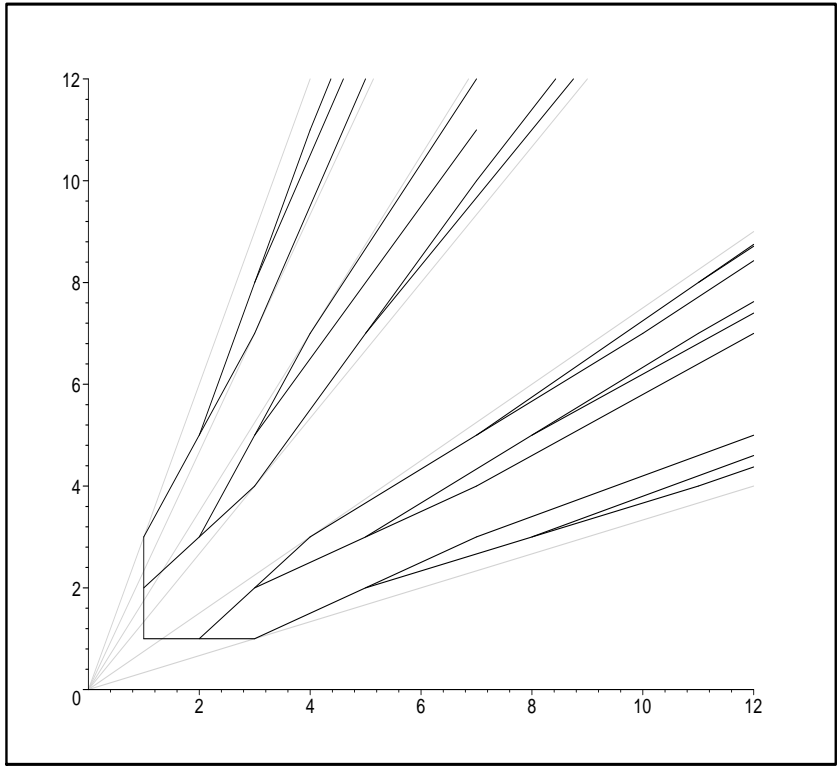}
\caption{L'arbre de r\'esolution ${\cal R}_3$.}
\end{figure}

Le principal effet de la contrainte de r\'esolution est d'ouvrir les z\^ones du plan asso\c{c}i\'ees \`a un rationnel donn\'e.
Par ailleurs, les nombres compris dans cette z\^ones sont envoy\'es par $r_a$ sur le rationnel correspondant au noeud.

\section{Construction dynamique et th\'eor\`eme de structure}

La construction pr\'ec\'edente sur $\Z^2$ donne une vision g\'eom\'etrique qui n'est pas adapt\'ee \`a une comparaison directe
avec le spectre de fr\'equences exp\'erimental. L'ensemble ${\cal R}_a$ peut se visualiser en tra\c{c}ant le graphe de
la fonction not\'ee $e_a :\R \rightarrow \R$ et d\'efinie par
\begin{equation}
x\longmapsto \mid x-r_a (x) \mid ,
\end{equation}
qui donne l'erreur d'approximation.\\

On obtient la figure suivante:

\begin{figure}[h]
\label{lap01}
\centering
\includegraphics[width=1\textwidth]{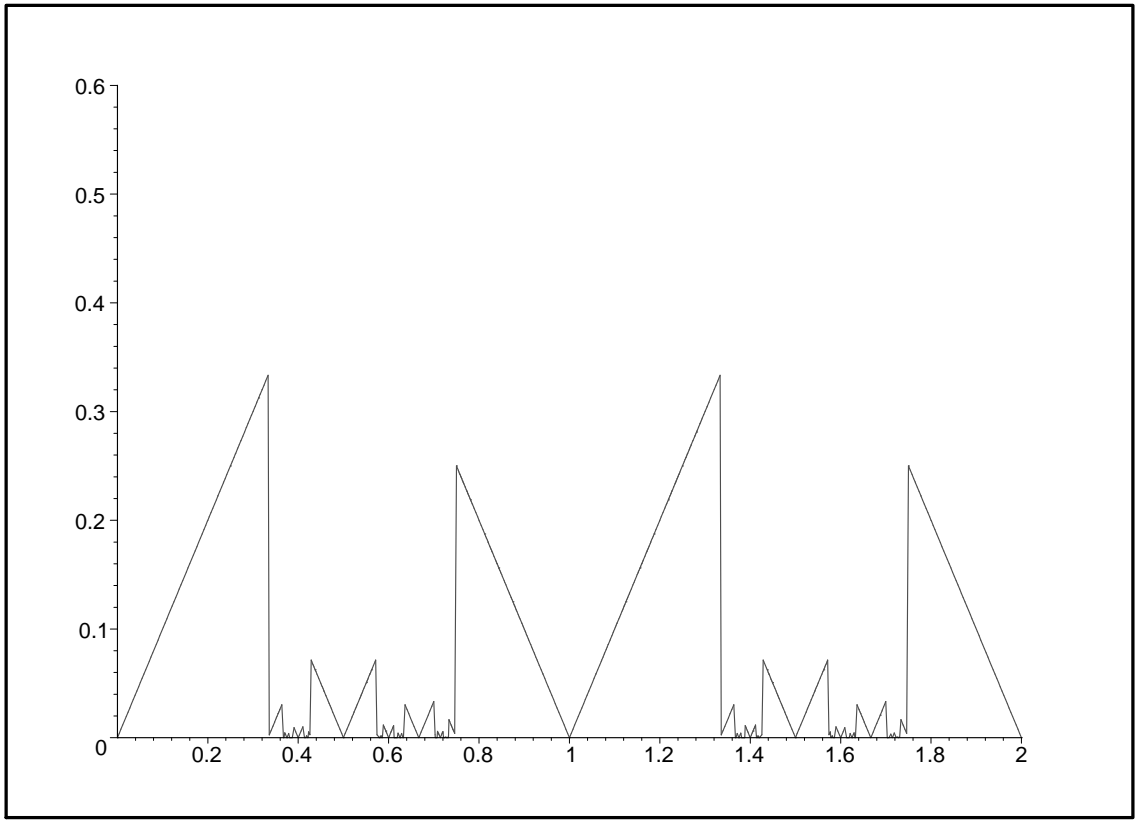}
\caption{Graphe de l'erreur d'approximation pour $a=3$}
\end{figure}

On peut aussi la faire ``\`a la main" de mani\`ere it\'erative, en transportant la premi\`ere structure qui apparait,
\`a savoir la z\^one d'accumulation au voisinage de z\'ero, et en regardant ce que devient cette structure via les
op\'erations $x\mapsto x+1$ et $x\mapsto 1/x$. Le tra\c{c}\'e de la fonction d'erreur d'approximation apparait ainsi
tout seul. Il montre aussi comment la premi\`ere z\^one d'accrochage apparait au voisinage de $1$ par transport
de la z\^one d'accumulation en $0$ via la translation et l'inversion laissant fixe le point $1$. Cette construction
\`a l'avantage d'\^etre simple et parlante.\\

Nous avons le th\'eor\`eme de structure suivant :

\begin{thm}
Soit $a\in \nN^*$, l'ensemble de r\'esolution ${\cal R}_a$ se
d\'ecompose en:
\pesp
i - rationnels attractifs : soit $p/q$ un tel rationnel, il d\'efini un interval
d'accrochage $I^a_{p/q} = [\nu^- (p/q) ,\nu^+ (p/q)]$ tel que pour tout $x\in I_{p/q}$,
on a $r_a (x)=p/q$.
\pesp
ii - rationnels transitoires : soit $p/q$ un tel rationnel, il d\'efini un
interval de transit \`a droite (resp. \`a gauche) $I^+_{p/q} =[p/q , \nu^+ (p/q)]$
(resp. $I_{p/q}^- =[\nu^- (p/q) ,p/q]$) tel que pour tout $x\in I^+ (p/q)$ (resp.
$x\in I^- (p/q)$), on a $r_a (x)=p/q$.
\pesp
iii - irrationnels de blocage : ils sont obtenus comme accumulation de zones
de blocage.
\pesp
iv - irrationnels transitoire : ils sont obtenus comme accumulation de zones de transit.
\pesp
v - irrationnels mixtes : soit $\xi$ un tel irrationnel. Il est obtenu comme accumulation de zones
de transit et de blocage.
\end{thm}

Ce th\'eor\`eme n'est qu'une retraduction du fait que l'ensemble de r\'esolution ${\cal R}_a$ est un arbre. On peut aussi
le voir directement via la construction it\'erative de ${\cal R}_a$.

\subsection{Sur les z\^ones d'accrochages}
\label{zoneaccro}

Dans ce paragraphe, on travaille dans un ensemble de r\'esolution donn\'e ${\cal R}_a$, $a\in \nN^*$.\\

Soit $x=p/q$ un rationnel de blocage. On a

\begin{lem}
\label{func}
Pour tout nombre rationnel de blocage $x=p/q\in {\cal R}_a$, on a
\begin{equation}
\left .
\begin{array}{lll}
i\ \ -\ \nu^{\sigma} (1+x) & = & 1+\nu^{\sigma} (x) , \ \ \ \sigma =\pm, \\
ii\ -\ \nu^{\sigma } (1/x) & = & \di {1\over \nu^{-\sigma } (x)} .
\end{array}
\right .
\end{equation}
\end{lem}

Ces relations gardent un sens pour tout nombre $x\in \rR^*$, ce qui
permettra de ne plus pr\'eciser si on travaille avec un rationnel de blocage.

\begin{lem}
Soit $p/q =[a_0 ,\dots ,a_n ] \in {\cal R}_a$, on a
\begin{equation}
\left .
\begin{array}{lll}
\nu^{\sigma } & = & [a_0 ,\dots ,a_n ,a] ,  \\
\nu^{-\sigma } & = & [a_0 ,\dots ,a_n -1 ,1,a ] ,
\end{array}
\right .
\end{equation}
avec $\sigma =+$ si $n$ est pair et $\sigma =-$ si $n$ est impair.
\end{lem}

\begin{proof}
On fait la d\'emonstration pour $\nu^+$, la
d\'emarche \'etant analogue pour $\nu^-$. On a
$$\nu^+ ([a_0 ,\dots ,a_n ]) =\nu^+ (a_0 +\di {1\over [a_1 ,\dots ,a_n ]} ) =a_0 +
\nu^+ (\di {1\over [a_1 ,\dots ,a_n ]} ) ,$$
par l'\'egalit\'e i) du lemme \ref{func}. De plus, on a
$$\nu^+ (\di {1\over [a_1 ,\dots ,a_n ]} )=\di {1\over \nu^- ([a_1 ,\dots ,a_n ] )} ,$$
par ii). Une simple r\'ecurrence donne donc
$$\nu^+ ([a_0 ,\dots ,a_n ])=[a_0 ,\dots ,a_{n-1} ,\nu^{\sigma } (a_n ) ],$$
avec $\sigma =+$ si $n$ est impair et $\sigma =-$ sinon.

Comme on a pour tout entier $0< m <a$, $\nu^+ (m)=m+\di {1\over a}$, $\nu^- (m) =m-1+\di {1\over 1+\di {1\over a}}$
et de plus, $\nu^+ (0)=1/a$, $\nu^- (a)=a-1+\di {1\over 1+\di {1\over a}}$, on en d\'eduit le lemme.
\end{proof}

Ce r\'esultat est le plus frappant vis \`a vis des donn\'ees exp\'erimentales. Les valeurs du bord des z\^ones d'accrochage pr\'edites
via ce lemme sont en accord quasi parfait avec celles obtenues exp\'erimentalement (voir \cite{serge} et \cite{mar}).

\subsection{Bassin d'attraction d'un rationnel}

Soit $a\in\N^*$ et $p/q$ un rationnel donn\'e de ${\cal R}_a$. Le bassin d'attraction de $p/q$, not\'e ${\cal A}(p/q)$,
est d\'efini comme
\begin{equation}
{\cal A} (p/q)=\left \{
x\in \R,\ \exists k\in\N,\ r_a^k (x)=p/q \right \} ,
\end{equation}
o\`u $r_a^k =r_a \circ \dots r_a$ $k$ fois.\\

Ces bassins sont form\'es de la z\^one d'accrochage proprement dire et des z\^ones transitoires accol\'ees.\\

Avant de donner une caract\'erisation du bord du bassin d'attraction, regardons un exemple ou $r_a$ agit non
trivialement:\\

Soit $a=3$ et $p/q=[0,1,2,1,3]$. On a $r_3 (p/q)=[0,1,2,1]=[0,1,3]$ et $r_3^2 (p/q)=[0,1]$.\\

On voit donc ici un exemple de dynamique des approximations via l'application $r_3$. Ce ph\'enom\`ene est d\^u \`a
l'existence de rationnels dont la fraction continue est de la forme
\begin{equation}
[a_1 ,\dots ,a_n ,a_1 ,1,a] .
\end{equation}
Pour ces nombres l'action de $r_a$ ne donne pas de suite la bonne approximation. En effet, on a
\begin{equation}
r_a ([a_1 ,\dots ,a_n ,a_1 ,1,a] )=[a_1 ,\dots ,a_n ,a] ,
\end{equation}
soit
\begin{equation}
r_a^2 ([a_1 ,\dots ,a_n ,a_1 ,1,a] )=[a_1 ,\dots ,a_n ] .
\end{equation}
L'\'evolution dynamique de l'approximation de $[a_1 ,\dots ,a_n ,a_1 ,1,a]$ s'arr\'ete si et seulement si $a_n <a$. Le
ph\'enom\`ene ci-dessus est \`a l'origine de la terminologie de rationnels {\it transitoires} dans le th\'eor\`eme de
structure.\\

Le lemme suivant caract\'erise simplement le bord du bassin d'attraction d'un rationnel:

\begin{lem}
Soit $a\in \N^*$ et $[a_1 ,\dots ,a_n]$ un rationnel donn\'e de ${\cal R}_a$. Les bords de son bassin d'attraction sont des
irrationnels quadratiques. Pr\'ecis\'ement, les valeurs des bords sont
\begin{equation}
[a_1 ,\dots ,a_n ,a-1 ,1,\dots , a-1 ,1,\dots ]  \ \mbox{\rm et}\
[a_1 ,\dots ,a_n -1,1, a-1 ,1,\dots , a-1 ,1,\dots ] .
\end{equation}
\end{lem}

\begin{proof}
La d\'emonstration repose sur la construction it\'erative du bord de la z\^one d'accrochage en $1$. On transporte ensuite
ces bords pour obtenir le rationnel choisi. Nous allons faire la construction pour le bord droit du bassin d'attraction, le bord
gauche n'offrant pas plus de difficult\'es.\\

Une z\^one transitoire \'etant donn\'ee \`a droite de $1$, on obtient la prochaine en appliquant les op\'erations suivantes:
$x\mapsto 1/x$, $x\mapsto x+a-1$, $x\mapsto 1/x$ er $x\mapsto x+1$. Autrement dit, on it\`ere l'application
\begin{equation}
t_a (x)=\di 1+{x\over 1+x(a-1)} .
\end{equation}
Les points fixes de cette fonction sont des irrationnels quadratiques. La forme de l'application $t_a$ traduite sur les
fractions continues nous dit que ces irrationnels s'obtiennent en collant aux fractions continues du bord des z\^ones
d'accrochage une suite infinie de $a-1,1$.
\end{proof}

On peut \'etudier d'autres types de nombres irrationnels obtenus comme par exemple accumulation de z\^ones de blocage.
On renvoie \`a (\cite{CrDe},p.317-318) pour un exemple. N\'eanmoins, ces r\'esultats sont difficiles \`a tester et interpr\'eter
au niveau exp\'erimental et physique.

\section{Approche dynamique du spectre des fr\'equences}

L'analyse pr\'ec\'edente permet une reconstruction globale du spectre des fr\'equences, mais ne dit pas la mani\`ere dont
les fr\'equences bougent au cours du temps lors de la d\'etection du signal. Or, de r\'ecentes exp\'eriences de Michel
Planat et Jean-Philippe Marillet \cite{mar} ont mis en \'evidence l'existence de sauts de fr\'equences au voisinage des
r\'esonances. Ce paragraphe donne une base th\'eorique \`a ce ph\'enom\`ene fond\'ee sur l'hypoth\`ese diophantienne.

\subsection{Dynamique des fractions continues}

Pour tout $\nu\in R\setminus \R$, on note $p_i /q_i$ son $i$-\`eme convergent. Pour chaque valeur de $i$, on regarde la z\^one d'accrochage
attach\'ee au rationnel $p_i /q_i$. La taille de cette z\^one est proportionnelle \`a $q_i$. Pour comprendre la dynamique
des fractions continues sous l'hypoth\`ese diophantienne on doit \'etudier l'\'evolution des $q_i$ lorsque $i$ croit.

\subsection{Exposants de stabilit\'e}

Il est possible de quantifier les variations de $q_i$ lorsque $i$ croit. Pour tout $i\geq 1$, il existe un unique r\'eel
$\tau_i \geq 1$ et $\gamma_i >0$ tel que
\begin{equation}
q_{i+1} =\gamma_i q_i^{\tau_i},
\end{equation}
avec $1\leq \gamma_i <q_i$. L'exposant $\tau_i$ peut se concevoir comme un exposant caract\'erisant la stabilit\'e de
$q_i$ lorsque $i$ croit.\\

L'analyse diophantienne fournit des renseignements int\'eressants sur cet exposant:

\begin{lem}
Soit $\nu\in \R$, ses exposants de stabilit\'e sont uniform\'ement born\'es si et seulement si $\nu$ est un nombre diophantien.
\end{lem}

Ce lemme d\'ecoule du th\'eor\`eme de Siegel \cite{sie}: un nombre r\'eel $\nu $ est diophantien si et seulement si il existe
$\gamma >0$ et $\tau \geq 1$ tels que $q_{i+1} \leq \gamma q_i^{\tau}$.\\

On peut aussi \'etudier l'\'evolution des $q_i$ via la {\it fonction de Brujno} \cite{br}:\\

La fonction de Brujno, not\'ee ${\cal B}$, est d\'efinie pour tout $\nu\in \R\setminus \Q$, par
\begin{equation}
{\cal B} (\nu )=\di\sum_{i\geq 0} \di {\log q_{i+1} \over q_i } ,
\end{equation}
o\`u $p_i/q_i$ est le $i$-\`eme convergent de $\nu$.\\

On renvoie au travail de S. Marmi, P. Moussa et J-C. Yoccoz \cite{mmy} pour une \'etude d\'etaill\'ee des propri\'et\'es de
cette fonction.

\subsection{Instabilit\'e au voisinage des r\'esonances}

Commen\c{c}ons par un fait exp\'erimental: lorsque $f_0 =1.00000007$ MHz et $f_1 =0.599975$ MHz, on a
\begin{equation}
\di {f_0 \over f_1} =0.599974958...=[0,1,1,2,1596,1,10,\dots ] .
\end{equation}
On observe que le d\'etecteur effectue des sauts autour des valeurs suivantes de fr\'equence de battement:
\begin{equation}
f=135,\ 261,\ 386\ {\rm Hz} .
\end{equation}
Comment comprendre ce ph\'enom\`ene ?\\

Pour tout $a\in\N^*$, on note $\nu (a)$ le nombre $[0,1,1,2,a]$ et $p(a)/q(a)$ son \'ecriture sous forme de fraction
irr\'eductible. Si on note $f(a)$ la fr\'equence d\'efinie par
\begin{equation}
f(a)=\mid p(a) f_0 -q(a) f_1 \mid ,
\end{equation}
on obtient pour $a=1593,\ 1594$ et $1595$ les fr\'equences de battement $135$, $261$ et $386$ respectivement.\\

Autrement dit, les sauts de fr\'equences observ\'ees correspondent \`a des fluctuations des quotients partiels, en particulier
du param\`etre de troncature.\\

On peut comprendre cette situation de la fa\c{c}on suivante: lorsque le d\'enominateur du $i$-th convergent devient instable
(i.e. lorsqu'on a une augmentation brusque du quotient partiel dans le d\'eveloppement en fraction continue), on a des
z\^ones d'accrochage tr\`es fines. De ce fait, le syst\`eme devient sensible aux perturbations.

\section{R\'ealit\'e ou artefact ?}

La th\'eorie que nous avons propos\'e n'explique pas pourquoi le syst\`eme fait de l'approximation diophantienne.
Il me semble que si une raison claire existe elle doit se trouver du cot\'e de la physique microscopique et d'une
compr\'ehension
plus fine de la physique des m\'elangeurs.\\

On peut aussi
mettre en doute le fait que les effets de hi\'erarchie que nous avons observ\'e sont dus au syst\`eme physique et donc
mettent en \'evidence finalement des propri\'et\'es de la nature. Cette suspicion tient au fait que nous n'avons pas
acc\'e \`a des donn\'ees bruts. En effet, entre l'exp\'erience proprement dite et les
donn\'ees se trouve un ordinateur pour l'acquisition et le traitement des donn\'ees (il y a une phase de comptage sur
le signal). Rien ne dit que la fa\c{c}on d'effectuer ce comptage et du m\^eme coup tout le traitement des donn\'ees n'est
pas finalement biais\'e. Cette situation est in\'evitable et entre en fait dans tout proc\'ed\'e de mesure d'un syst\`eme
physique.\\

Je ne crois pas qu'il soit possible de trancher pour le moment. Il me semble que le probl\`eme est de m\^eme nature que celui de d\'ecider
si le monde r\'eel est un continuum ou discret. Je renvoie \`a la discussion de E. Schr\"odinger (\cite{scr},p.41-59) pour plus
de d\'etails.\\

Ce qui est s\^ur c'est que de nombreux probl\`emes de physiques font intervenir d'une mani\`ere ou d'une autre des
r\'esolutions, i.e. des limites \`a notre mesure du r\'eel. Cette limitation n'a dans certains cas que peut d'incidence,
comme dans l'\'etude de beaucoup de ph\'enom\`enes macroscopiques. La pr\'ecision toujours plus grande des mesures,
notamment dans le cas des oscillateurs, nous fait toucher du doigt il me semble la structure infime du r\'eel. On tombe
alors sur des ph\'enom\`enes nouveaux mais de port\'ee universelle. Je renvoie encore une fois au texte de E. Schr\"odinger
(\cite{scr},p.49-59) ou, sur quelques pages, il donne une construction tr\`es proche dans l'esprit des espaces de
r\'esolutions pour d\'emontrer les difficult\'es li\'ees \`a l'hypoth\`ese d'une nature continue.

\end{document}